\documentclass[letterpaper,12pt]{article}   
\usepackage{osajnl2} 
\usepackage{hyperref} 
\usepackage{graphicx}
\usepackage{subfigure}

\begin{document}

\title{Bright Chirp-free and Chirped Nonautonomous solitons under Dispersion and Nonlinearity Management}


\author{Zhan-Ying Yang,$^{1,*}$ Li-Chen Zhao,$^1$ Tao Zhang,$^1$ Rui-Hong Yue$^2$}

\address{$^1$Department of  Physics, Northwest University, Xi'an
  710069, China}
\address{$^2$Faculty of Science, Ningbo University, Ningbo, 315211, China}
\address{$^*$Corresponding author: zyyang@nwu.edu.cn}

\begin{abstract}We present a series of chirp-free and
  chirped analytical nonautonomous soliton solutions to the
  generalized nonlinear Schr\"{o}dinger equation (NLSE) with
  distributed coefficients by Darboux transformation from a
  trivial seed. For chirp-free nonautonomous soliton, the
  dispersion management term can change the motion of
  nonautonomous soliton and do not affect its shape at
  all. Especially, the classical optical soliton can be
  presented with variable dispersion term and nonlinearity
  when there is no gain. For chirped nonautonomous soliton,
  dispersion management can affect the shape and motion of
  nonautonomous solitons meanwhile. The periodic dispersion
  term can be used to control its ``breathing'' shape, and it
  does not affect the trajectory of nonautonomous soliton
  center with a certain condition.

\end{abstract}


\maketitle 
\section{Introduction}
It is well known that ideal optical soliton in fiber, theoretically
reported by Hasegawa and Tappert \cite{Hasegawa01,Hasegawa02} and experimentally verified by
Molenauer et al.\cite{Mollenauer:1980zz}, is based on the exact balance between the group
velocity dispersion and the self-phase modulation. However, it is
very difficult to realize ideal optical soliton  communication due
to  fiber loss. The dissipation would weaken the nonlinearity and
finally the optical soliton would broaden and lose its signal\cite{3}.
There are  two ways to overcome these inferiors. One
 is to compensate for the fiber loss by optical gain via Raman
amplification \cite{4}. The other  is to use  dispersion management and
nonlinearity management, which have been investigated in recent
years \cite{5} and \cite{6}. When the two techniques are performed, the
dynamics of  optical pulse propagation will be governed by
 inhomogeneous nonlinear Schr\"{o}dinger equation
(INLSE)
\begin{equation}\label{eq1}
i\frac{\partial U}{\partial Z}+\Omega(Z) \frac{\partial^2
U}{\partial T^2}+ R(Z)|U|^2U+\frac{iG(Z)}{2}U=0\ ,
\end{equation}
where $U(Z,T)$ is the complex envelope of an electrical field in a
co-moving frame, $Z$ and $T$ are  normalized distance and retarded
time. The coupling parameters $\Omega(Z)$, $R(Z)$ and $G(Z)$ are
free function, standing for the group velocity of dispersion
management, Kerr nonlinearity and gain parameter respectively. It is
clear that these parameters all have affected the soliton
solution\cite{Serkin01,Serkin02,Serkin03,Kruglov,9,10}. Based on this property, the concept of soliton
management has been proposed with the development of modern
technology \cite{Malomed}, which essentially is to control the soliton's
dynamics by tuning the related parameters. since then, how to
control exactly the soliton's dynamics by dispersion and
nonlinearity management with gain term becomes an issue worth
considering. Luo et al. have presented some ways to manage them\cite{3}.
Here, we propose  a direct way to control the soliton's dynamics by
tuning the related controllable parameters. Meanwhile, in the frame
of new management the dynamical behavior of optical soliton can be
studied more exactly and conveniently through the explicit
expressions of nonautonomous soliton's peak, width and the
trajectory of envelope's center. Especially, these expressions
provide a potential application to the design of fiber optic
amplifiers, optical pulse compressors, and solitary wave- based
communication links.

In this article, we
suggest two different ways to manage the dispersion and nonlinearity
with gain based on the exact chirp-free and chirped soliton
solutions of the Eq.(1). For the chirp-free nonautonomous soliton,
the dispersion just affects soliton's motion without changing its
shape. For the chirped soliton, the dispersion management can affect
both its shape and trajectory of the wave center with the certain
condition. The explicit expressions in the two circumstances are
presented, which describe the shape and trajectory of nonautonomous
soliton. Excitingly, the precise condition for classical soliton
under dispersion management is achieved as $R(Z)=2g\Omega(Z)$
($g\neq0$). The classical soliton under periodic dispersion
management can be presented with no gain for chirp-free soliton,
whose center just oscillate periodically, and its shape keep
invariant. Oppositely, there is a kind of chirped soliton under the
same management which just "breath" and its center does not
oscillate under a certain condition.

\section{ Bright Chirp-free Nonautonomous soliton Solution }

To get an analytical solution of the INLS equation, there should be
some constrains (integrable conditions) on the coupling
parameters\cite{3,Serkin01,Serkin02,Serkin03,Kruglov,9,10}. These conditions show a subtle balance among the
dispersion term, gain(loss) and nonlinearity, which has a profound
implication to control the soliton's dynamics. Here, we choose
\begin{eqnarray}
R(Z)&=&2g\Omega(Z)\exp {[\int G(Z)d Z]}\ ,\label{condition1}\\
U(\tau,Z)&=&Q(T,Z)\exp{[-\int G(Z)/2 dZ]}\ ,
\end{eqnarray}
then the Eq.(1) changes into
\begin{equation}
iQ_Z+\Omega(Z)Q_{TT}+f(Z)T Q+2g\Omega(Z)|Q|^2Q=0
\end{equation}
with the Lax-Pair
\begin{equation}
\partial_T \phi=F\phi(T,Z),\quad \partial_Z\phi =W\phi(T,Z)
\end{equation}
\begin{equation}
F=\left(
\begin{array}{cc}
  \zeta &\sqrt{g}Q\\
  -\sqrt{g}\bar{Q}&-\zeta \\
\end{array}
\right) ,  \quad  W=\left(
\begin{array}{cc}
 A &B\\
  C &-A \\
\end{array}
\right).
\end{equation}
Here
\begin{eqnarray}
A&=&2i\Omega(Z)\zeta^2+ig\Omega(Z)|Q|^2,\nonumber\\
B&=&2i\sqrt{g}\Omega(Z)Q\zeta+i\sqrt{g}\Omega(Z)Q_T,\nonumber\\
C&=&-2i\sqrt{g}\Omega(Z)\bar{Q}\zeta
+i\sqrt{g}\Omega(Z)\bar{Q}_T,\nonumber
\end{eqnarray}
and $\zeta=\alpha+i\beta$  an arbitrary complex number. By
performing Darboux transformation from a trivial seed, one can find
an analytic solution of INSE with condition (\ref{condition1})
\begin{equation}
U[T,Z]=\frac{4\alpha A_c \exp\theta}{\sqrt{g}[1+A_c^2\exp\varphi]}\
,
\end{equation}
where
\begin{eqnarray}
\theta&=&-2T(\alpha-i\beta)-\int\!dZ\,G(Z)/2\nonumber\\
 &+&i4(\alpha-i\beta)^2\int\! dZ\Omega(Z)\nonumber\\
\varphi&=& 16\alpha \beta \int\! dZ\Omega(Z) -4\alpha T ,\nonumber
\end{eqnarray}
and $A_c$ an arbitrary real number.

Note that in the above solution, the coupling parameters $\Omega(Z)$
and $G(Z)$ are assumed as  arbitrary $Z$-dependent functions, which
will be very convenient to study the properties of nonautonomous
solitons with each different conditions. And the solution is  a
chirp-free nonautonomous soliton. It is easy to find the chirp
parameter to be zero from the definition of chirp parameter.

As usual, the position of the maximum value is defined as  the
center of envelope, which is given by $1-A_c^2e^{\varphi}=0$ in
present case. Thus, the wave center of the nonautonomous soliton is
given by:
\begin{equation}\label{tc}
T_c=\frac{\ln A_c} {2\alpha} +4\beta \int \Omega(Z)d Z
\end{equation}
Obviously,  the dispersion management term dominates the trajectory
of the wave center effectively while the gain parameter has no
effect on nonautonomous soliton's trajectory.

The evolution of bright nonautonomous soliton's width defined  by
the half-value corresponding width can be  given as following
\begin{equation}\label{width1}
W(Z)=\frac{1}{4\alpha} \ln \frac{3+2\sqrt{2}}{3-2\sqrt{2}},
\end{equation}
and the evolution of soliton's peak can be described by the function
\begin{equation}
|U|_{max}^2=4 \alpha^2 \exp[-\int G(Z)d Z]/|g|.
\end{equation}
The equation (\ref{width1}) reflects a fact that the coupling parameters appeared 
in equation (\ref{eq1}) do not affect the
evolution of nonautonomous soliton's width. For this
 soliton, the dispersion management just affects the
trajectory of soliton's wave center, while the gain parameter
governs the peak.  Thus, one can study the properties of solitons
with many kinds of dispersion through the general solution. Besides,
if there is no gain, the classical soliton will be recovered with
the condition $R(Z)=2g\Omega(Z)$($g\neq 0$).

 Since the exponentially dispersion
management has been realized experimentally\cite{11,12,13}, we choose the
dispersion management as $\Omega(Z)=e^Z$. Then the solution which
describes the evolution of nonautonomous soliton in the waveguide
can be presented as
\begin{equation}
|U|^2=\frac{16\alpha^2 A_c^2 \exp[\varphi-\int G(Z)d
Z]}{g[1+A_c^2\exp\varphi]^2}
\end{equation}
where $\varphi=16\alpha \beta e^Z-4\alpha T$.

From the expression (8), we  know that the nonautonomous soliton will depart
from the propagation direction for $\Omega(Z)=e^{Z}$. If
$\Omega(Z)=e^{-Z}$, it will approach propagation direction.
Especially, if $G(Z)=0$ both  soliton width and peak don't change at
all, which is a classical optical soliton, and has great potential
application in the fibers communication system\cite{Hasegawa01}.

It is well known that the periodic dispersion management has been
achieved for a long time\cite{15,16,17}. Here we observe the evolution of
nonautonomous soliton with the dispersion management term
$\Omega(Z)=l cos(\omega Z)$. From this case,  the soliton's center
is oscillating with propagation distance. So one can get the
oscillating soliton under periodic dispersion management in the
retarded frame, shown in Fig.1. We stress that the evolutions of
nonautonomous solitons under dispersion management are all observed
in the retarded frame. It should be noted that this oscillating
soliton is quite different from the so-called dispersion-managed
soliton which is always chirped and its shape is changed\cite{18}. This
means that the dispersion-managed soliton in present paper is
chirp-free  whose shape is invariable from the balance of dispersion
and nonlinear effects.

\section{CHIRPED NONAUTONOMOUS SOLITON SOLUTION}
 If the nonlinearity is
chosen as $R(Z)=2g \Omega(Z)\exp{[\int
 (G(Z)-4\Omega(Z)C_2(Z))dZ]}$, where the function $C_2(Z)=1/(\int 4\Omega(Z)d
 Z+h)$ ($h$ is a constant),
 the nonautonomous soliton solution can be given as following by performing Darboux
transformation method
\begin{equation}
U(T,Z)=\frac {4 b(Z) A_c \exp[\theta(T,Z)]}{\sqrt{g}(1+A_c^2
\exp[\varphi(T,Z)])}
\end{equation}
where (with three arbitrary real numbers $\alpha$, $\beta$ and
$A_c$)
\begin{eqnarray}
\theta(T,Z)&=&-2[b(Z)-id(Z)]T+iC_2(Z)T^2\nonumber\\
&+&\int[-G(Z)/2+2\Omega(Z)C_2(Z)]d Z\nonumber\\
&+&\int 4i \Omega(Z)[b(Z)-id(Z)]^2 d Z,\nonumber\\
\varphi(T,Z)&=&-4b(Z) T+\int 16\Omega(Z)b(Z)d(Z)d Z ,\nonumber\\
b(Z)&=&\alpha \exp[-\int 4\Omega(Z) C_2(Z)d Z],\nonumber\\
d(Z)&=&\beta \exp[\int -4\Omega(Z) C_2(Z) d Z].\nonumber
\end{eqnarray}


From the definition of chirp parameter, one can find  the chirp
parameter to  be $2C_2(Z)$. So chirped nonautonomous soliton
solution is achieved. The other quantities  charactering the
soliton's shape and trajectory can be calculated as following.

The evolution of width is
\begin{equation}
W(Z)=\frac{1}{4 b(Z)} \ln \frac{3+2\sqrt{2}}{3-2\sqrt{2}}.
\end{equation}
The evolution of its peak is
\begin{equation}
|U|_{max}^2=\frac{4 b(Z)^2}{g} \exp[\int 4\Omega(Z)
C_2(Z)-G(Z)dZ].
\end{equation}
And the trajectory of its center is
\begin{equation}
T_c(Z)=\frac{\ln A_c}{2b(Z)} +\frac{\int 4\Omega(Z) b(Z) d(Z)
dZ}{b(Z)}.
\end{equation}
From the above equations, we  know that the dispersion term affects
both the trajectory and the shape for chirped soliton solution,
which is quite different from the chirp-free one. In order to
compare with a chirp-free nonautonomous soliton in last section, we
also take   periodic dispersion management $\Omega(Z)=l\cos(\omega
Z)$ without the gain term. The nonlinearity parameter becomes
$R(Z)=\frac{gl\omega \cos(\omega Z)}{4l\sin(\omega Z)+\omega h}$,
and the corresponding chirp parameter is $\frac{\omega}{2l \sin
 (\omega Z)+\omega h/2}$ ($|h|>4|l|/\omega$),
the trajectory of its center can be given from Eq.(15) as
\begin{equation}
T_c(Z)=\frac{(4l\sin(\omega Z)/\omega+h)\ln A_c}{2\alpha} -\beta.
\end{equation}

It is clear that the coefficient $A_c$ has a critical physical
effect. For $A_c=1$, the trajectory of soliton's center will becomes
$T_c(Z)=-\beta$, which means that the soliton's center does not
oscillate any more. The trajectory of wave center is a straight
line. Moreover, from Eq.(13) and (14), one can find that the
soliton's width evolves as
\begin{equation}
W(Z)=\frac{4l\sin(\omega Z)+\omega h}{4\alpha \omega}\ln
\frac{3+2\sqrt{2}}{3-2\sqrt{2}},
\end{equation}
and the peak is
\begin{equation}
|U|_{max}^2=\frac{4\alpha^2 \omega}{g[4l\sin(\omega Z)+\omega h]}.
\end{equation}
From these expressions, the soliton is obviously a "breather" whose
shape is changed periodically, shown in Fig.2. Both the width and
the peak of  nonautonomous soliton are oscillating. This is quite
different from the chirp-free nonautonomous soliton under the same
dispersion management, which just changes chirp-free nonautonomous
soliton's group velocity, but keeps the shape invariant. This  can
also  be seen by comparing Fig.1 and Fig.2(a). Additionally, to show
the character of "breathing soliton", we draw the contour plot of
Fig.2(a) in Fig.2(b). For $A_c\neq 1$, it breathes and its
group velocity changes periodically such as Fig.3. Interestingly ,
chirped nonautonomous soliton with $A_c\neq 1$ has characters of
both chirp-free one and chirped one with $A_c=1$. So, the soliton
under periodic dispersion management can evolve as oscillating
soliton, or breathing soliton, or oscillating breathing soliton
depending on  different conditions. This  is to be constructive to
control the soliton.

If dispersion management is chosen as $l\cos(\omega Z)+l_0$, we can
also get the evolution of nonautonomous soliton from Eq.(12). When
$l_0>0$, the soliton will breath more and more lightly with the
propagation distance, shown in Fig.4(a); when $l_0<0$, it will
breath more and more heavily, such as Fig.4(b). This an interesting
feature of the soliton in present paper. The above discussions are
all made with vanishing gain term. Similarly, one can discuss no
vanishing gain term case and find  that the gain term does not
affect the width and motion of soliton, but changes  the peak of
soliton. From the Eq.(14), one  can find that the soliton's peak is
 constant when the gain term becomes $G(Z)=-4\Omega(Z) C_2(Z)$.

\section{DISCUSSION AND CONCLUSION}
In summary, we have deduced the exact bright chirp-free and chirped
nonautonomous soliton solutions of the nonlinear equation with
arbitrary $Z$-dependent coupling parameters. The functions of such
parameters can be seen clearly through the equations  describing the
evolution of  soliton's center, width and peak. An advantage of
present scenario is  convenience,  one can  study the evolution of
nonautonomous soliton in various kinds of systems. If the
integrability condition is $R=2g\Omega(Z)\exp [\int G(Z)d Z]$,
chirp-free soliton can be achieved. The freedom of choosing
dispersion management can be used to control the motion of soliton,
which does not affect its shape at all with the corresponding
nonlinearity. And the gain parameter just affects the peak of the
soliton. This provides a lot of particular ways to control the
evolution of soliton in waveguide amplifiers or optical fibers.
Moreover, for chirp-free
 solitons without the gain, the classical soliton with
unchanged shape can be presented with the condition that the
 dispersion term $\Omega(Z)$ and nonlinear
parameter  are related by $R(Z)$ is $R(Z)=2g\Omega(Z)$($g\neq 0$),
which is helpful to improve the quality of soliton transmission.
 For periodic dispersion management
$\Omega(Z)=l\cos(\omega Z)$,  a novel soliton solution exists, and
the center just oscillate and its shape does not change at all.

If the integrability condition  $R(Z)= 2g \Omega(Z)\exp{[\int
 G(Z)-4\Omega(Z)C_2(Z)dZ]}$ with $C_2(Z)
=\frac{1}{\int 4\Omega(Z)d Z+h}$ is satisfied, the chirped
nonautonomous soliton can be achieved. Especially, the chirped
 soliton under the same dispersion management
$\Omega(Z)=l\cos(\omega Z)$ have been studied. When the gain $G(Z)$
vanishes, there is a novel "breathing" soliton  with the condition
 $R(Z)=\frac{gl\omega \cos(\omega Z)}{4l\sin(\omega Z)+\omega
h}$ and  $A_c=1$, whose wave center did not oscillate but the shape
changed. If $A_c\neq1$,  the chirped soliton' center and its shape
will both change under the same dispersion management.

It is worth to point out that the chirp-free nonautonomous soliton
in present paper keeps its shape and trajectory unchanged, while the
dispersion affect both the trajectory and shape of  the chirped
nonautonomous soliton. This is quite different  form the known
dispersion-managed solitons. It provides a possible application in
designing some optic apparatus such as amplifier,  pulse compressors
and etc.

\textbf{Acknowledgement:} This paper was supported by Natural Science 
Foundation of China granted by No 10975180 and No 10875060.

\clearpage

\section*{List of Figure Captions}

Fig. 1. (Color online) The dynamics of chirp-free bright temporal nonautonomous
soliton under periodic dispersion management for  $\alpha = 2, l =
1, \omega = 2, g = 0.25, A_c = 2, \beta= 1$.

Fig. 2. (Color online) (a) The dynamics of chirped bright nonautonomous soliton
under periodic dispersion management for $\alpha = 2; l = 3; \omega
= 3; g = 0.25; A_c = 1; \beta = 1; h=5$. (b) The contour plot of (a)
with the same parameters. It is obviously that the soliton is
"breathing", and both width and peak oscillate with time. 

Fig. 3. (Color online) (a) The dynamics of chirped bright nonautonomous solitons
 under periodic dispersion management and $A_c = 8, \alpha = 2, l = 3, \omega =
 3, g = 0.25, \beta = 1, h=5$ . (b) The contour plot  with  same
parameters. It is obviously that the soliton is "breathing",its
width and peak oscillate with time, and its center oscillates too. 

Fig. 4. (Color online) The dynamics of bright solitons under dispersion
management $\Omega (Z)=l\cos(\omega Z)+l_0$. (a) $l_0=0.03$,the
nonautonomous soliton breathes more and more lightly; (b)
$l_0=-0.03$, it breathes more and more heavily. The other
coefficients are: $\alpha =2; l=3; \omega =3; A_c=1; g=0.25;
\beta =1; h=5$.

\clearpage

\newpage
\begin{figure}
\centering
\includegraphics[width=40mm,height=40mm]{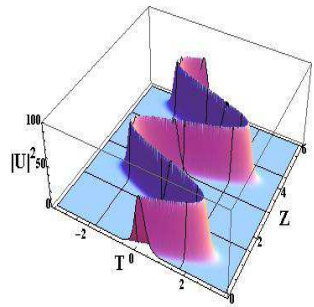}
\caption{(Color online) The dynamics of chirp-free bright temporal nonautonomous
soliton under periodic dispersion management for  $\alpha = 2, l =
1, \omega = 2, g = 0.25, A_c = 2, \beta= 1$.}
\end{figure}

\clearpage
\begin{figure}
\centering
\subfigure[]{\includegraphics[height=40mm,width=40mm]{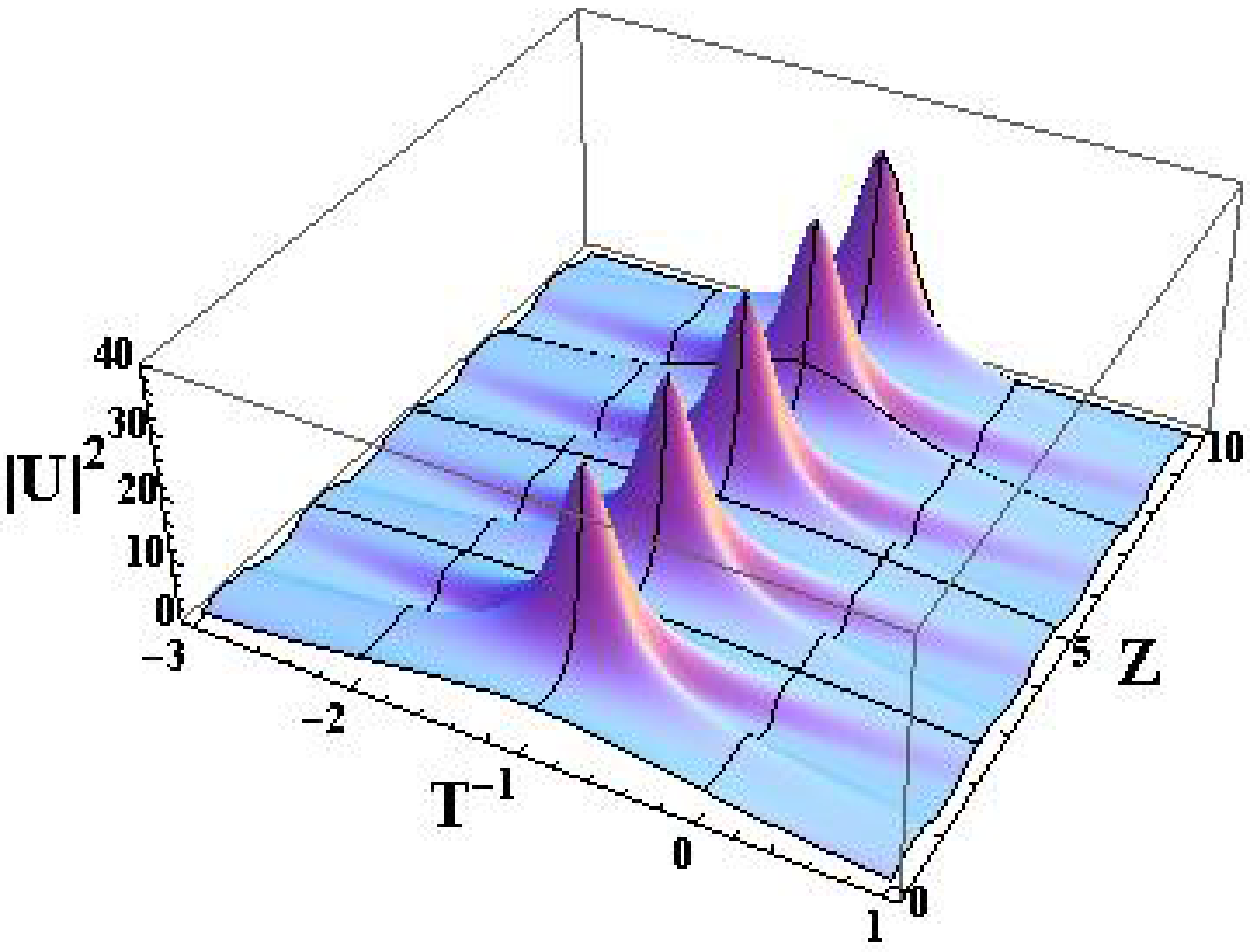}}
\hfil
\subfigure[]{\includegraphics[height=40mm,width=40mm]{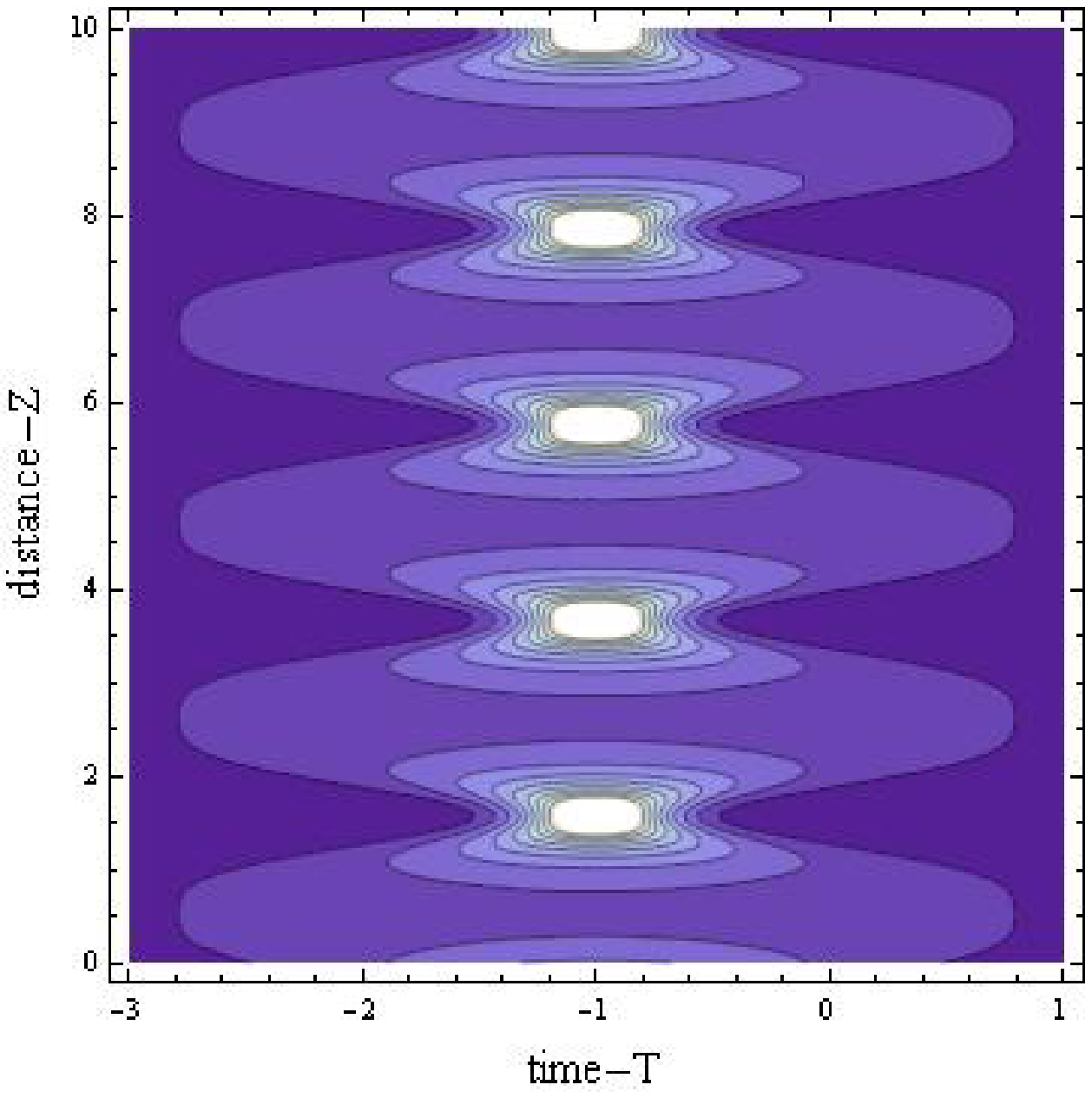}}
\caption{(Color online) (a) The dynamics of chirped bright nonautonomous soliton
under periodic dispersion management for $\alpha = 2; l = 3; \omega
= 3; g = 0.25; A_c = 1; \beta = 1; h=5$. (b) The contour plot of (a)
with the same parameters. It is obviously that the soliton is
"breathing", and both width and peak oscillate with time. }
\end{figure}

\clearpage
\begin{figure}
\centering
\subfigure[]{\includegraphics[height=40mm,width=40mm]{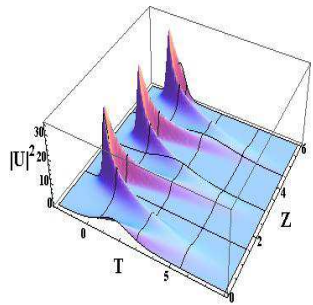}}
\hfil
\subfigure[]{\includegraphics[height=40mm,width=40mm]{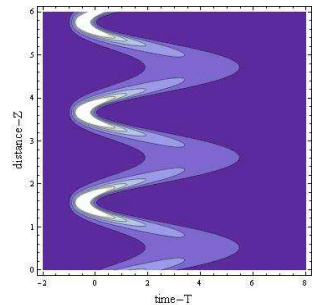}}
\caption{(Color online) (a) The dynamics of chirped bright nonautonomous solitons
 under periodic dispersion management and $A_c = 8, \alpha = 2, l = 3, \omega =
 3, g = 0.25, \beta = 1, h=5$ . (b) The contour plot  with  same
parameters. It is obviously that the soliton is "breathing",its
width and peak oscillate with time, and its center oscillates too. }
\end{figure}

\clearpage
\begin{figure}
\centering
\subfigure[]{\includegraphics[height=40mm,width=40mm]{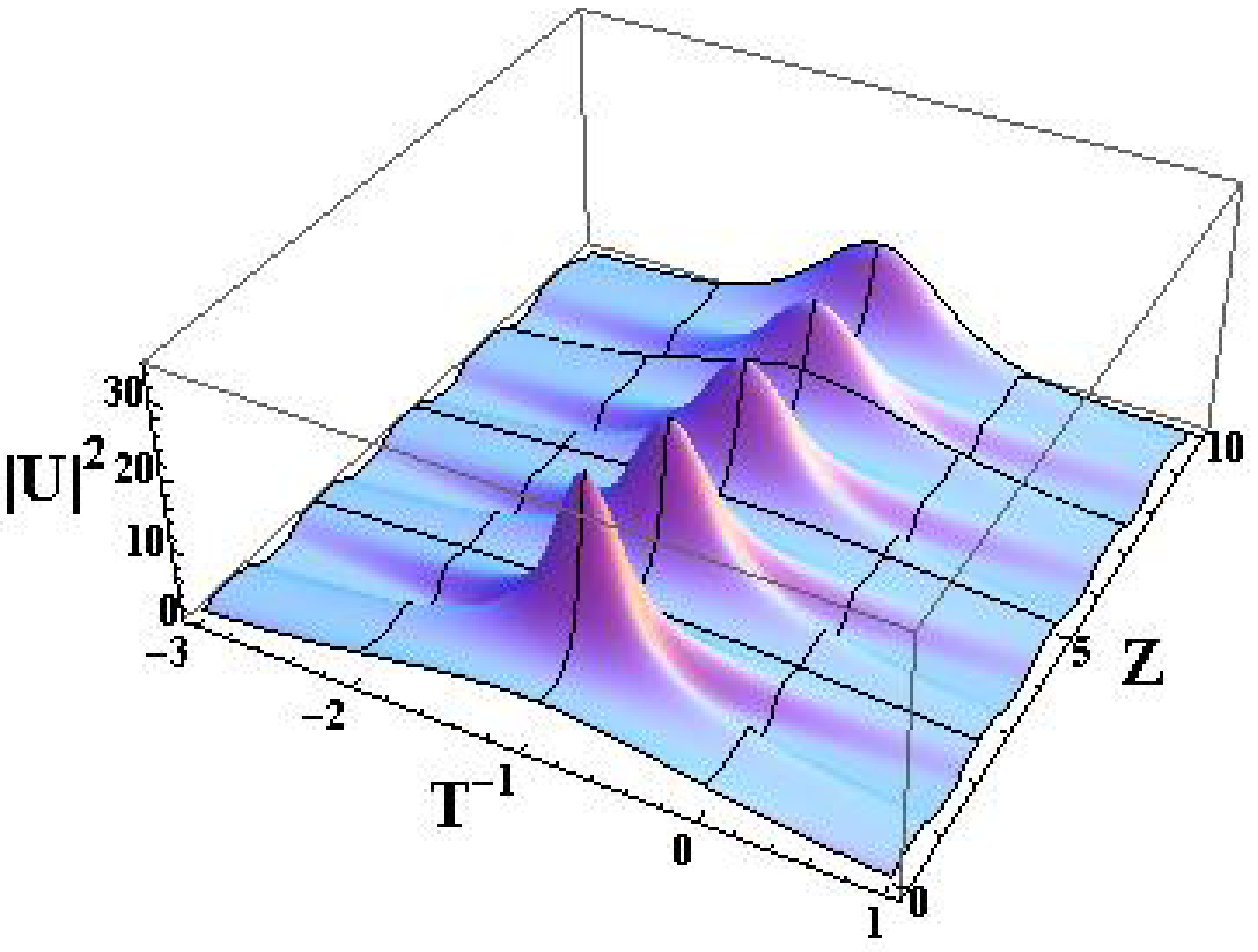}}
\hfil
\subfigure[]{\includegraphics[height=40mm,width=40mm]{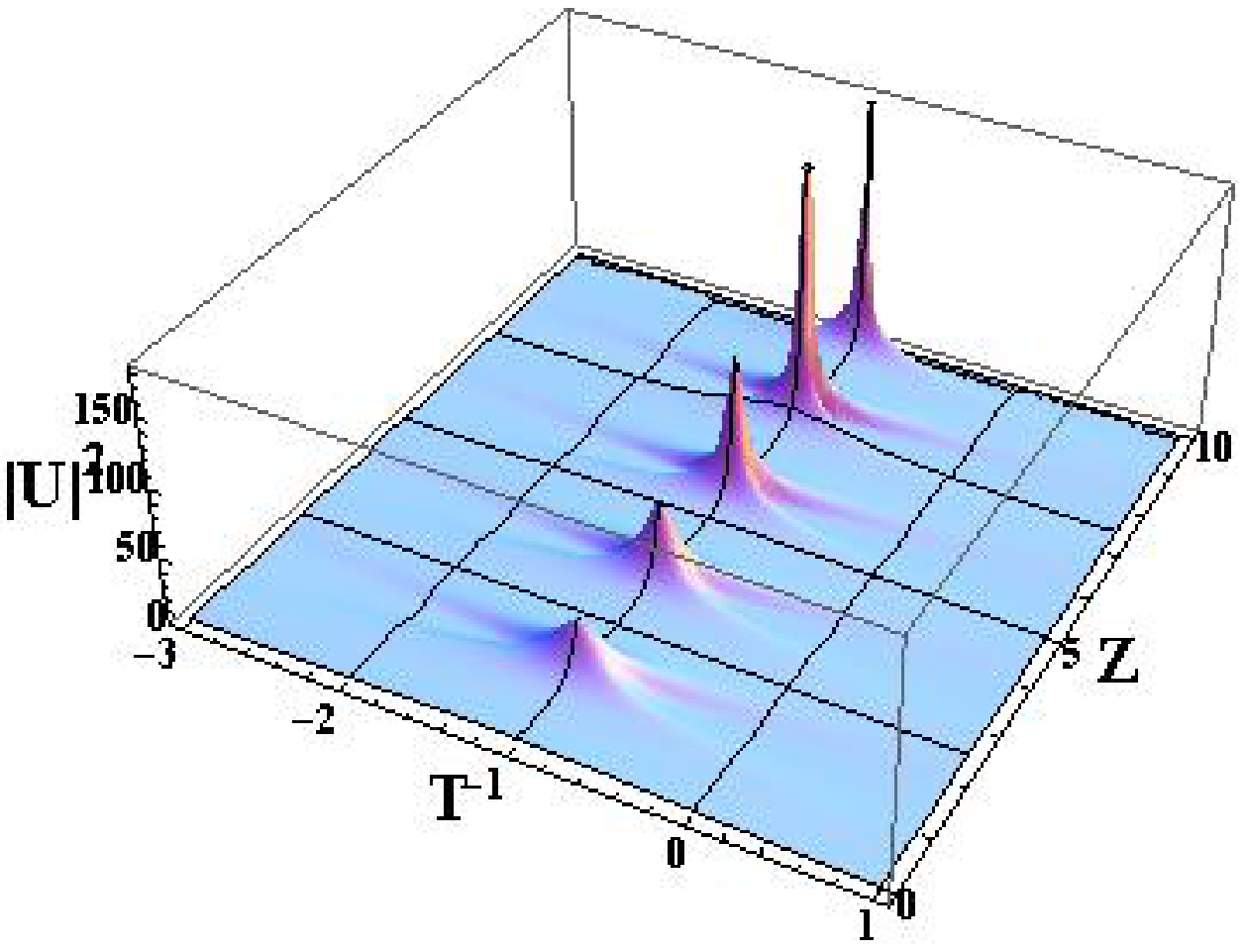}}
\caption{ (Color online) The dynamics of bright solitons under dispersion
management $\Omega (Z)=l\cos(\omega Z)+l_0$. (a) $l_0=0.03$,the
nonautonomous soliton breathes more and more lightly; (b)
$l_0=-0.03$, it breathes more and more heavily. The other
coefficients are: $\alpha =2; l=3; \omega =3; A_c=1; g=0.25;
\beta =1; h=5$.}
\end{figure}
\end{document}